\documentclass{article}
\usepackage[T1]{fontenc}
\usepackage[utf8]{inputenc}
\usepackage{physics}
\usepackage{xcolor}
\usepackage{amsmath}
\usepackage{tikz}
\usepackage{booktabs}
\usepackage{slashed}
\usepackage{amssymb}
\usepackage{bm}
\usepackage{graphicx}
\usepackage[justification=centering]{caption}
\usepackage[left=4cm,right=4cm,top=2cm,bottom=2cm]{geometry}
\usepackage[numbers]{natbib}
\usepackage{hyperref}
\usepackage{authblk}
\hypersetup{
    colorlinks=true,
    linkcolor=red,
    filecolor=magenta,      
    urlcolor= magenta,
    citecolor=blue
}
\usepackage{url}
\usepackage[nottoc]{tocbibind}
\usepackage{appendix}
\date{}

\title{Effective flavor interaction for the charged-lepton mass hierarchy and the Koide relation}
\author{Olivier Rousselle\thanks{\href{mailto:olivierrousselle@live.fr}{\texttt{rousselle@zitter-institute.org}}}}
\affil{Fondation Louis de Broglie, 23 rue Marsoulan, 75012 Paris, France \\ Sorbonne Université, 4 place Jussieu, 75005 Paris, France}

\begin{document}
\maketitle
\begin{abstract}
This article proposes a dynamical framework to address the charged-lepton mass hierarchy. We postulate a $U(3)_F$ flavor symmetry at the high-energy scale, where a dynamical flavon field couples to the electroweak Higgs sector via an effective operator. As the universe cools, spontaneous symmetry breaking of the flavor group occurs following the pattern $U(3)_F \to U(1)^3_F$. The effective potential governing the flavon vacuum relaxes into a stable minimum, enforcing an algebraic equipartition between the flavor-singlet and flavor-octet invariants ($I_{(0)} = I_{(8)}$). Subsequently, the electroweak symmetry breaking at the Fermi scale generates the charged-lepton mass matrix, successfully identifying the eigenvalues with the masses of the electron, muon, and tau. 
\end{abstract}

\section{Introduction}

Understanding the origin of the fermion mass spectrum and flavor mixing remains one of the central open problems of the Standard Model of particle physics. While the gauge sector is controlled by only five free parameters, the Yukawa sector contains twelve fermion masses and at least eight mixing parameters \cite{goffinet2005}. Since the advent of the Standard Model, physicists have aimed at reducing these numbers. The Standard Model thus requires some new flavor physics, to explain the fermion mass spectrum and the number of generations \cite{goffinet2005}.
A striking hint in this direction was discovered by Y. Koide in 1981, i.e. the pole masses of the charged leptons (electron, muon, tau) satisfy the empirical equation \cite{koide1981}:
\begin{equation}
Q=\frac{m_e+m_\mu+m_\tau}{\left(\sqrt{m_e}+\sqrt{m_\mu}+\sqrt{m_\tau}\right)^2}=\frac{2}{3}
\end{equation}
with remarkable precision. Using the 2026 PDG values \cite{PDG2026}  $m_e \approx 0.51099895069(16)~\mathrm{MeV}/c^2$, $m_\mu \approx 105.6583755(23)~\mathrm{MeV}/c^2$ and $m_\tau\approx 1776.93(09)~\mathrm{MeV}/c^2$, we get $Q=0.66666446(508)$, a deviation from 2/3 of order $2\times 10^{-6}$, well within experimental uncertainties. 
The ratio $Q$ is necessarily bounded between 1/3 and 1, with the lower bound $Q = 1/3$ corresponding to fully degenerate masses, and the upper limit $Q \to 1$ characterizing an extreme hierarchy. The Koide ratio is invariant under any permutation of the mass parameters $(m_e,m_\mu,m_\tau)$, suggesting a deep geometrical connection between the three fermion generations.

We formalize and extend the theoretical frameworks pioneered by Koide \cite{koide1990} and Sumino \cite{sumino2009} to address the charged-lepton mass hierarchy. This is achieved by introducing several key theoretical components: a fundamental global $U(3)_F$ flavor symmetry; a flavor sector featuring a dynamical flavon field distinct from the electroweak Higgs; an ultraviolet (UV) completion that yields the effective Yukawa interaction; and a full spontaneous symmetry breaking pattern $U(3)_F \to U(1)_F^3$ driven by an effective potential whose minimization enforces an equipartition condition. The complete dynamical mechanism is detailed in Section \ref{sec:yukawa}, while the resulting charged-lepton mass spectrum is analyzed in Section \ref{sec:leptons}.

\section{Effective Yukawa interaction and mass generation} 
\label{sec:yukawa} 

\subsection{Dynamic flavon field}

In the Standard Model (SM), the Yukawa sector describes the coupling between the Higgs doublet $H$ and the three generations of fermion fields \cite{weinberg1967}: $\mathcal{L}_Y^\textrm{SM} = -Y \overline{\Psi}_L H \Psi_R + \text{h.c.}$, where $\Psi_L$ and $\Psi_R$ are the left- and right-handed chiral fermions, and $Y$ is the $3\times 3$ Yukawa matrix. We postulate $U(3)$ as the flavor symmetry, recognizing it as the exact global symmetry of the kinetic Lagrangian for massless fermions in the SM. 

To address the origin of flavor dynamically, we extend the theory beyond the electroweak scale by introducing a heavy, scalar, dynamical flavon field $\Phi$ (of mass dimension 1). The flavon is a Hermitian $3\times 3$ matrix-valued field:
\begin{equation}
\Phi(x) = \sum_{a=0}^8 \phi_a(x)\,T_a
\end{equation}
where $\phi_a$ are real scalar fields, $T_0=\frac{I_3}{\sqrt{6}}$, and $T_a=\frac{\lambda_a}{2}$ ($a=1,\dots,8$) are the Hermitian generators of $SU(3)$, with $\lambda_a$ being the Gell-Mann matrices. 

The massless fermion fields $\Psi_L$ and $\Psi_R$ transform as fundamental triplets $\mathbf{3}$ under $U(3)_F$, and the flavon field transforms in the adjoint representation of $U(3)_F$: $\Phi \to U \Phi U^\dagger$, where $U \in U(3)_F$. The Lagrangian terms associated with the flavon sector are:
\begin{equation}
\mathcal{L} = \mathcal{L}_\textrm{kin}(\Phi) - V_F(\Phi) + \mathcal{L}_Y(\Phi, H, \Psi_L, \Psi_R)
\end{equation} 
where $\mathcal{L}_\textrm{kin}(\Phi)=\Tr\left[\partial_\mu\Phi \partial^\mu\Phi\right]$ is the canonical kinetic term, $V_F(\Phi)$ is the flavon potential, and $\mathcal{L}_Y$ is the $U(3)_F$-invariant Yukawa interaction whose effective expression at energy much below UV scale $\Lambda_F$ is:
\begin{equation}
\mathcal{L}_Y^\textrm{eff} = -\frac{c_F}{\Lambda_F^2} \overline{\Psi}_L \Phi\, H\,\Phi\,\Psi_R + \text{h.c.} = -\frac{c_F}{\Lambda_F^2} (\overline{\Psi}_L)_{ib} \,\Phi^i_j\,H^b\,\Phi^j_k (\Psi_R)^k + \text{h.c.}
\end{equation} 
where $i,j,k \in \{1,2,3\}$ are the flavor indices, $b \in \{1,2\}$ is the $SU(2)_L$ gauge index, and $c_F>0$ is a dimensionless Wilson coefficient. The relevant quantum numbers for the charged-lepton sector are summarized in Table~\ref{tab:reps}.
\begin{table}[ht]
  \centering
  \begin{tabular}{@{}lccc@{}}
    \toprule
    Field & $SU(2)_L$ & $U(1)_Y$ & $U(3)_F$ \\
    \midrule
    $\Psi_L$ & $\bm 2$ & $-1/2$ & $\bm 3$ \\
    $\Psi_R$ & $\bm 1$ & $-1$ & $\bm 3$ \\
    $H$ & $\bm 2$ & $+1/2$ & $\bm 1$ \\
    $\Phi$ & $\bm 1$ & $0$ & $\bm 1\oplus\bm 8$ \\
    \bottomrule
  \end{tabular}
  \caption{Representations in the charged-lepton sector.}
  \label{tab:reps}
\end{table}  \\
Because the Higgs field is a flavor singlet, it commutes with the flavon matrix in flavor space, allowing the quadratic insertion $\Phi^2$ to effectively play the role of the dynamical Yukawa matrix $Y$. 
This specific dimension-6 effective operator $\mathcal{L}_Y^\textrm{eff}$ was introduced by Sumino \cite{sumino2009} to build a theoretical framework for Koide's mass formula. Physically, the flavon field acts as a dynamical bridge connecting the fundamental chiral fermions to the electroweak Higgs sector, with one flavon acting on $\Psi_L$ and the other on $\Psi_R$. 

This effective interaction is realized via an ultraviolet (UV) completion inspired by the Froggatt-Nielsen mechanism \cite{froggatt1979}. By introducing heavy vector-like fermionic messengers $F_1$/$F_2$ with bare masses $M_1$/$M_2$ and generating the chain $\Psi_L\xrightarrow{\Phi}F_1\xrightarrow{H}F_2\xrightarrow{\Phi}\Psi_R$, the mass and interaction terms of the UV Lagrangian read:
\begin{align}
\mathcal{L}_Y^{\mathrm{UV}} = - M_1 \overline{F}_1 F_1 - M_2 \overline{F}_2 F_2 - \Bigl( y_1 \overline{\Psi}_L \Phi F_{1R} + y_2 \overline{F}_{1L} H F_{2R} + y_3 \overline{F}_{2L} \Phi \Psi_R + \text{h.c.} \Bigr).
\label{eq:UVLag}
\end{align}
Integrating out these heavy vector-like fields at energies well below the $M_{1,2}$ mass scale yields the effective interaction $\mathcal{L}_Y^{\mathrm{eff}}$, mapping the UV parameters to the effective theory with $c_F=y_1y_2y_3$ and $\Lambda_F^2=M_1M_2$.

\subsection{Symmetry breaking mechanism}

\paragraph{Flavor symmetry breaking}

The Hermitian flavon field $\Phi$ can be decomposed orthogonally into $\Phi = \Phi_{(0)} + \Phi_{(8)}$, where $\Phi_{(0)} = \frac{1}{3}\Tr[\Phi]I_3$ represents the $U(1)$ flavor-singlet component, and $\Phi_{(8)} = \Phi - \frac{1}{3}\Tr[\Phi]I_3$ is the traceless adjoint $SU(3)$ flavor-octet component. 
We can introduce the two independent bilinear invariants for these components: 
\begin{equation}
I_{(0)}\equiv \Tr\left[\Phi_{(0)}^2\right]=\frac{1}{2}\phi_0^2\quad,\quad I_{(8)}\equiv \Tr\left[\Phi_{(8)}^2\right]=\frac{1}{2}\sum_{a=1}^8\phi_a^2
\end{equation}
such that $\Tr\left[\Phi^2\right]=I_{(0)}+I_{(8)}$.
We can parameterize the tree-level flavon potential in terms of these algebraic invariants:
\begin{equation}
V_F = -\mu_F^2\left(I_{(0)} + I_{(8)}\right) + \frac{\kappa_F}{2} \left(I_{(0)} + I_{(8)}\right)^2 + \frac{\kappa'_F}{2} \left(I_{(0)} - I_{(8)}\right)^2
\label{tree-level-potential}
\end{equation}
with $\mu_F^2>0$, $\kappa_F>0$ and $\kappa'_F>0$. By recalling that $I_{(0)} = \frac{1}{2}\phi_0^2$ and $I_{(8)} = \frac{1}{2}\sum_{a=1}^8 \phi_a^2$, it becomes clear that this algebraic structure effectively introduces a direct biquadratic coupling between the singlet amplitude and the collective octet amplitude. This algebraic structure adopts the analog Ginzburg-Landau form encountered in two-component superconductivity \cite{silaev2012}. 

When the universe cools down below the cutoff scale $\Lambda_F$, it evolves from the unstable origin (where $\Phi=0$) via spontaneous symmetry breaking (SSB). At the flavor breaking scale $v_F<\Lambda_F$, the SSB of $U(3)_F$ generates a vacuum expectation value (VEV) for the flavon field, noted $\langle\Phi\rangle$. 
To find the true physical vacuum, we must minimize the potential $V_F$. The invariants $I_{(0)}$ and $I_{(8)}$ serve as coordinates on the space of configurations; as $\frac{\partial V_F}{\partial \phi_0}=\phi_0 \frac{\partial V_F}{\partial I_{(0)}}$ and $\frac{\partial V_F}{\partial \phi_a}=\phi_a \frac{\partial V_F}{\partial I_{(8)}}$, minimization within the invariants is equivalent to radial minimization with respect to the scalar fields. Taking the partial derivatives of $V_F$ with respect to those invariants, we obtain the following system of stationary equations:
\begin{eqnarray}
\frac{\partial V_F}{\partial I_{(0)}} = -\mu_F^2 + \kappa_F\left(I_{(0)}+I_{(8)}\right)+\kappa_F'\left(I_{(0)}-I_{(8)}\right) = 0 \\
\frac{\partial V_F}{\partial I_{(8)}} = -\mu_F^2 + \kappa_F\left(I_{(0)}+I_{(8)}\right)-\kappa_F'\left(I_{(0)}-I_{(8)}\right) = 0
\end{eqnarray}
This yields the equipartition:
\begin{equation}
I_{(0)} = I_{(8)}=\frac{\mu_F^2}{2\kappa_F}\equiv\frac{v_F^2}{4}.
\end{equation}
such that $\Tr[\langle\Phi\rangle^2]=\frac{v_F^2}{2}$. An analysis of the Hessian matrix reveals that this symmetric equipartition configuration is a stable global minimum as $\kappa_F>0$ and $\kappa'_F>0$.
Moreover, since the flavon VEV $\langle\Phi\rangle$ is a Hermitian matrix, the spectral theorem allows us to use the global $U(3)_F$ symmetry to rotate its VEV into a basis where $\langle \Phi \rangle$ is real and diagonal:
\begin{equation}
\langle \Phi \rangle = \frac{v_F}{\sqrt{2}}\,\text{diag}(\lambda_1,\lambda_2,\lambda_3),
\end{equation}
with $\lambda_k>0$ and $\sum_{k=1}^3\lambda_k^2=1$. Assuming non-degenerate eigenvalues, this SSB follows the pattern $U(3)_F \to U(1)^3$. 
According to Goldstone's theorem \cite{goldstone1962}, since the fundamental group $U(3)_F$ has 9 generators and the residual group $U(1)^3$ has 3 generators, this spontaneous breaking dictates the emergence of $9-3=6$ massless Nambu-Goldstone bosons. Moreover, the flavon multiplet contains two massive radial excitations: a symmetric radial mode with tree-level mass $m_{h_R}=\sqrt{\kappa_F}v_F=\sqrt{2}\mu_F$, and a singlet–octet alignment mode with mass $m_{h_A}=\sqrt{\kappa_F'}v_F$.

Following the standard eigenvalue parametrization for a $3\times 3$ flavor space, any set of three real numbers $(\lambda_1,\lambda_2,\lambda_3)$ can be parametrized as:
\begin{equation}
\lambda_k = a_0 + 2\,a_8\cos\left(\theta_F + \frac{2k\pi}{3}\right),
\end{equation}
with $a_0>0$, $a_8>0$ and $\theta_F \in [0,2\pi)$. The normalization condition $\sum_{k=1}^3\lambda_k^2=1$ leads to $3\,a_0^2+6\,a_8^2=1$.
Finally, the equipartition condition $I_{(0)}=I_{(8)}$, i.e. $\Tr\left[\langle\Phi\rangle_{(0)}^2\right] = \Tr\left[\langle\Phi\rangle_{(8)}^2\right]$, translates to:
\begin{equation}
\Tr\left[\langle\Phi\rangle^2\right]=\frac{2}{3}\left(\Tr[\langle\Phi\rangle]\right)^2 \iff \sum_{k=1}^3\lambda_k^2 = \frac{2}{3}\left(\sum_{k=1}^3\lambda_k\right)^2 \iff a_0=\sqrt{2}\,a_8.
\end{equation}

\paragraph{Electroweak symmetry breaking and generation of the masses}

At the Fermi scale $v \approx 246.22$ GeV, the Electroweak Spontaneous Symmetry Breaking (EWSB) occurs via the Brout-Englert-Higgs mechanism \cite{higgs1964, englert1964}. The Higgs field acquires its vacuum expectation value $\langle H \rangle = (0, v/\sqrt{2})^\text{T}$. The overall symmetry breaking pattern of the physical vacuum concludes as:
\begin{equation}
U(3)_F \times SU(2)_L \times U(1)_Y \longrightarrow U(1)^3_F \times SU(2)_L \times U(1)_Y \longrightarrow U(1)^3_F \times U(1)_\textrm{em}
\end{equation}
The effective Lagrangian that generates the tree-level masses is:
\begin{eqnarray}
\mathcal{L}^\textrm{eff}_\textrm{mass} &=& -\frac{c_F}{\Lambda_F^2} \bar{\Psi}_L \langle\Phi\rangle \,\langle H\rangle \, \langle\Phi\rangle \, \Psi_R + \textrm{h.c.} \notag \\
&=& -\sum_{k=1}^3\frac{v}{\sqrt{2}}\,\left(\frac{\sqrt{c_F}\,v_F}{\sqrt{2}\,\Lambda_F}\right)^2\lambda_k^2\,\bar{\Psi}^{k,2}_L\,\Psi^k_R  + \textrm{h.c.} \notag \\
&=& -\sum_{k=1}^3 m_k \bar{\Psi}_k\Psi_k
\end{eqnarray}
where $\Psi^{k,2}_L$ is the lower (charged) component of the left-handed $SU(2)_L$ fermion doublet, and where the left- and right-handed components combine into Dirac fermions $\Psi_k=\Psi_L^{k,2}+\Psi_R^k\in\mathbb{C}^4$. For charged leptons, $\Psi=(\Psi_e, \Psi_\mu, \Psi_\tau)^\textrm{T}$.
The resulting tree-level masses are given by:
\begin{equation}
m_k = \frac{v}{\sqrt{2}}\left(\frac{\sqrt{c_F}\,v_F}{\sqrt{2}\,\Lambda_F}\right)^2\lambda_k^2.
\label{masses}
\end{equation}
The mass matrix is proportional to the effective Yukawa matrix, which is determined by the flavon VEV:
\begin{equation}
M = \frac{v}{\sqrt{2}}Y_\textrm{eff}\quad\textrm{with} ~ Y_\textrm{eff}=\frac{c_F}{\Lambda_F^2} \langle\Phi\rangle^2.
\end{equation}
As $m_k \propto \lambda_k^2$, the constraint $\sum_{k=1}^3\lambda_k^2 = \frac{2}{3}\left(\sum_{k=1}^3\lambda_k\right)^2$ automatically leads to the Koide relation at the tree-level:
\begin{equation}
Q = \frac{\sum_k m_k}{\left(\sum_k \sqrt{m_k}\right)^2} = \frac{2}{3}.
\end{equation}

\paragraph{Analogy between the electroweak and flavor sectors}
To properly contextualize the proposed mechanism, we summarize in Table \ref{tab:ew_flavor_analogy} the structural parallels between the standard electroweak theory and our flavor framework. The Higgs boson $H$ is responsible for the electroweak symmetry breaking required for the masses of light leptons; whereas the flavon field $\Phi$ is the source of their flavor structure and hierarchy.

\begin{table}[h!]
    \centering
    \renewcommand{\arraystretch}{1.3} 
    \begin{tabular}{@{}lcc@{}}
        \toprule
        & \textbf{Electroweak sector} & \textbf{Flavor sector} \\
        \midrule
        Fundamental sym. & $SU(2)_L \times U(1)_Y$ (gauge) & $U(3)_F$ (global) \\
        Field & Higgs doublet $H\sim(\mathbf{2},1/2)$ & Flavon matrix $\Phi\sim\mathbf{1}\oplus\mathbf{8}$ \\
        Invariants & $X=H^\dag H$ & $I_{(0)}=\Tr[\Phi_{(0)}^2]$, $I_{(8)}=\Tr[\Phi_{(8)}^2]$ \\
        Potential & $V(X)=-\mu^2 X+\lambda X^2$ & $V_F\left(I_{(0)},I_{(8)}\right)$ \\
        Breaking scale & Fermi scale $v \approx 246$ GeV & UV Flavor scale $v_F \gg v$ \\
        VEV & $\langle H\rangle= \frac{1}{\sqrt{2}} \begin{pmatrix}0\\v\end{pmatrix}$ & $\langle\Phi\rangle= \frac{v_F}{\sqrt{2}}\text{diag}(\lambda_1,\lambda_2,\lambda_3)$ \\
        Residual sym. & $U(1)_\text{em}$ & $U(1)_F^3$ \\
        Angle & Weinberg angle $\theta_W$ & Flavor phase $\theta_F$ \\
        \bottomrule
    \end{tabular}
    \caption{Analogies between the standard Electroweak symmetry breaking mechanism and the proposed Flavor symmetry breaking mechanism.}
    \label{tab:ew_flavor_analogy}
\end{table}

\paragraph{Mass scale hierarchy}
The hierarchy between the electroweak scale and the lighter charged-lepton masses naturally arises from the ultraviolet suppression inherent to this framework. 
For fundamental UV couplings of order unity (i.e. $c_F \sim \mathcal{O}(1)$) and a hierarchy $v_F\ll\Lambda_F$, the charged-lepton masses are therefore suppressed relative to the electroweak scale by the squared ratio $(v_F/\Lambda_F)^2\ll1$. The overall suppression and the intergenerational hierarchy have distinct origins: the former is controlled by $v_F/\Lambda_F$, whereas the latter is encoded in the flavon eigenvalues through $m_k\propto\lambda_k^2$.

\section{Analysis of the charged-lepton mass spectrum}
\label{sec:leptons}

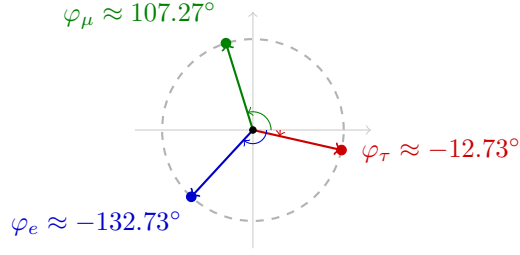
\begin{figure}[h!]
    \centering
    \begin{tikzpicture}[scale=1.2]
        \draw[thick, dashed, gray!60] (0,0) circle (1);
        \draw[->, gray!40] (-1.3,0) -- (1.3,0) node[right, text=black] {};
        \draw[->, gray!40] (0,-1.3) -- (0,1.3) node[above, text=black] {};
        \def\thetatau{12.73}
        \def\thetamu{-107.27}
        \def\thetae{132.73}
        \draw[->, thick, red!80!black] (0,0) -- (\thetatau:1);  \filldraw[red!80!black] (\thetatau:1) circle (1.5pt) node[right, xshift=4pt] {$\varphi_\tau \approx 12.73^\circ$}; 
        \draw[->, thick, green!50!black] (0,0) -- (\thetamu:1);  \filldraw[green!50!black] (\thetamu:1) circle (1.5pt) node[above left, yshift=-10pt] {$\varphi_\mu \approx -107.27^\circ$};    
        \draw[->, thick, blue!80!black] (0,0) -- (\thetae:1);        \filldraw[blue!80!black] (\thetae:1) circle (1.5pt) node[below left, yshift=10pt] {$\varphi_e \approx 132.73^\circ$};    
        \draw[->, red!80!black, thin] (0.3,0) arc (0:\thetatau:0.3) node[midway, right] {};
        \draw[->, green!50!black, thin] (0.2,0) arc (0:\thetamu:0.2) node[midway, right] {};
        \draw[->, blue!80!black, thin] (0.15,0) arc (0:\thetae:0.15) node[midway, right] {};
        \filldraw (0,0) circle (1pt);
    \end{tikzpicture}
    \caption{Angular representation of the cyclic charged-lepton mass parametrization.}
    \label{fig:Z3_leptons}
\end{figure}

The electron, the muon, and the tau emerge as the three fundamental nodes of a discrete $\mathbb{Z}_3$ cyclic network, and their physical masses share a common origin. 
Combining the normalization condition $3a_0^2+6a_8^2=1$ with the equipartition condition $a_0=\sqrt{2}\,a_8$, both derived in the previous section, fixes the two amplitudes: $a_0=\frac{1}{\sqrt{6}}$ and $a_8=\frac{1}{\sqrt{12}}$. Using those relations and the formula (\ref{masses}), we get this expression for the masses:
\begin{equation}
m_k=\frac{v}{\sqrt{2}}\epsilon_F^2\left(1+\sqrt{2}\cos\left( \theta_F+\frac{2k\pi}{3} \right) \right)^2,~k=1,2,3
\end{equation}
with $\epsilon_F=\frac{\sqrt{c_F}\,v_F}{\sqrt{12}\,\Lambda_F}$ the effective scaling parameter.
We can compute the flavor phase $\theta_F$ from the ratios of the experimental pole masses:
\begin{eqnarray}
\frac{\left(1+\sqrt{2}\cos(\theta_F-2\pi/3)\right)^2}{\left(1+\sqrt{2}\cos(\theta_F+2\pi/3)\right)^2}=\frac{m_\mu}{m_e}  \longrightarrow \theta_F \approx 0.22222~\textrm{rad}, \notag \\
\frac{\left(1+\sqrt{2}\cos\theta_F\right)^2}{\left(1+\sqrt{2}\cos(\theta_F+2\pi/3)\right)^2}=\frac{m_\tau}{m_e} \longrightarrow \theta_F \approx 0.22222~\textrm{rad}, \notag
\end{eqnarray}
indicating the consistency of the theory. The angles associated with the electron, muon and tau are:
$\varphi_e=\theta_F+\frac{2\pi}{3}\approx 132.73^\circ,~ \varphi_\mu=\theta_F-\frac{2\pi}{3}\approx -107.27^\circ,~ \varphi_\tau=\theta_F\approx 12.73^\circ$. A null mass is obtained for $\varphi=\pm 135^\circ$ and a maximum mass for $\varphi=0^\circ$. See Figure \ref{fig:Z3_leptons}. Moreover, the scaling parameter obtained is $\epsilon_F \approx 0.0425\ll1$.
 \\
The phenomenological flavor angle extracted from the experimental masses $\theta_F \approx 0.2222$ matches the rational fraction $2/9$ radians. This suggests that the vacuum orientation may encode an underlying geometric structure associated with the cyclic flavor sector. The dynamical origin of this specific phase $\theta_F$ is beyond the scope of this paper. 
In the electroweak sector, the on-shell renormalization scheme defines the mixing Weinberg angle $\theta_W$ directly from the physical pole masses of the W and Z bosons: $\sin^2\theta_W = 1-\frac{M_W^2}{M_Z^2}$. In an analogous manner, within our flavor sector, the angle $\theta_F$ is the geometric mixing angle that structures the charged-lepton pole masses. 

\paragraph{Koide relation for pole masses}
Following electroweak symmetry breaking, the effective Lagrangian generates the charged-lepton mass matrix. The flavor-vacuum alignment implies the Koide relation for the corresponding tree-level masses. These masses cannot, however, be identified directly with the physical pole masses, since QED self-energies and flavon loops generally introduce generation-dependent corrections. The remarkable accuracy of the empirical pole-mass relation therefore requires an additional radiative-protection or cancellation mechanism, which lies beyond the analysis presented here. As an illustration of this radiative sensitivity, the MS running charged-lepton masses at energy scale $\mu = M_Z$ give $Q(M_Z) \approx 0.6679$ \cite{xing2008}, rather than 2/3. This discrepancy illustrates the scale- and scheme-dependence of the relation constructed from renormalized masses.

\section{Conclusion}

The generation of charged-lepton masses in this model proceeds through a well-defined cascading sequence of symmetry breaking steps across distinct energy scales. The framework initiates at the ultraviolet cutoff scale $\Lambda_F$, where a global $U(3)_F$ flavor symmetry is preserved. As the universe cools to the intermediate flavor scale $v_F$ (with $v_F \ll \Lambda_F$), this symmetry undergoes spontaneous breaking, generating a vacuum expectation value for the flavon field. The system relaxes into a stable minimum characterized by a strict algebraic equipartition between the singlet and octet invariants ($I_{(0)} = I_{(8)}$). Further down at the electroweak Fermi scale $v$ (with $v \ll v_F$), the Higgs mechanism maps this geometric flavon VEV directly into the charged-lepton mass matrix. This dynamical mechanism successfully explains the charged-lepton mass hierarchy (with empirical Koide relation at tree-level) and the smallness of charged-lepton masses relative to the electroweak scale. While the Standard Model requires three independent and arbitrary Yukawa couplings, in our framework the charged-lepton spectrum is described by two parameters: one overall scaling factor $\epsilon_F$ and one phenomenologically extracted angle $\theta_F$. 

The apparent deviation from the Koide relation in the quark sector can be understood within this paradigm. Unlike leptons, quarks are strongly coupled to the $SU(3)_c$ color vacuum and are never observed as asymptotically free particles due to QCD confinement. Extending our geometric construction to the quark sector would therefore necessitate a scheme- and scale-dependent analysis involving strong running masses and non-perturbative QCD corrections, which lies beyond the scope of the present work. Furthermore, integrating the neutrino sector - potentially through a seesaw mechanism where right-handed neutrinos interact with the flavor vacuum under a distinct structural alignment - provides a compelling avenue for future investigation.

\paragraph{On the three generations} The fundamental triplication of the charged-lepton sector, parameterized in our model by a $U(3)_F$ flavor symmetry, could hypothetically stem from the maximal algebraic structure compatible with quantum mechanics. By enforcing probability conservation, Hurwitz's theorem limits the division rings to the octonions, where the requirement for observables to form a valid Jordan algebra strictly bounds the framework to the rank-3 Albert algebra $J_3(\mathbb{O})$ \cite{dubois2016, furey2016}. This exceptional geometric architecture inherently yields exactly three orthogonal projectors, providing a profound mathematical justification for the strict existence of exactly three fermion generations. 

\appendix

\section*{Acknowledgements}
The author warmly thanks Dominique Girardot for his insightful feedback and valuable intellectual exchanges, which helped shape the reflections presented in this manuscript.

\bibliographystyle{unsrt}
\bibliography{main}

\end{document}